\documentclass[]{IEEEphot}

\jvol{xx}
\jnum{xx}
\jmonth{June}
\pubyear{2009}

\usepackage{romannum}
\usepackage{amsmath}
\DeclareMathOperator{\sinc}{sinc}
\usepackage{graphicx}
\usepackage{multirow}
\usepackage{color}
\usepackage{ulem}

\begin{document}

\title{Wavelength Conversion Efficiency Enhancement in Modal Phase Matched $\chi^{(2)}$ Nonlinear Waveguides}

\author{Dongpeng Kang,$^1$~\IEEEmembership{Member,~IEEE}, Weiqi Zhang,$^1$ Amr S. Helmy,$^2$~\IEEEmembership{Senior~Member,~IEEE}, Siyuan Yu,$^1$ Liying Tan,$^1$ and Jing Ma$^{1}$}

\affil{1. School of Astronautics and National Key Laboratory of Science and Technology on Tunable Laser, Harbin Institute of Technology, 92 West Dazhi Street, Harbin, 150001, China\\
2. The Edward S. Rogers Department of Electrical and Computer Engineering, Centre for Quantum Information and Quantum Control, University of Toronto, 10 King's College Road, Toronto, Ontario M5S 3G4, Canada}

\doiinfo{DOI: 10.1109/JPHOT.2009.XXXXXXX\\
1943-0655/\$25.00 \copyright 2009 IEEE}%

\maketitle

\markboth{IEEE Photonics Journal}{Wavelength Conversion Efficiency Enhancement}

\begin{receivedinfo}%
Manuscript received March 3, 2008; revised November 10, 2008. First published December 10, 2008. Current version published February 25, 2009. This research was sponsored by the National Natural Science Foundation of China (61705053), China Postdoctoral Science Foundation (2016M600249), Heilongjiang Postdoctoral Special Funds, Fundamental Research Funds for the Central Universities, and Natural Sciences and Engineering Research Council of Canada.  Corresponding authors: Dongpeng Kang, Jing Ma (e-mail: dongpeng.kang@hit.edu.cn; majing@hit.edu.cn)
\end{receivedinfo}

\begin{abstract}
Modal phase matching (MPM) is a widely used phase matching technique in Al$_x$Ga$_{1-x}$As and other $\chi^{(2)}$ nonlinear waveguides for efficient wavelength conversions. The use of a non-fundamental spatial mode compensates the material dispersion but also reduces the spatial overlap of the three interacting waves and therefore limits the conversion efficiency. In this work, we develop a technique to increase the nonlinear overlap by modifying the material nonlinearity, instead of the traditional method of optimizing the modal field profiles. This could eliminate the limiting factor of low spatial overlap inherent to MPM and significantly enhance the conversion efficiency. Among the design examples provided, this technique could increase the conversion efficiency by a factor of up to $\sim$290 in an Al$_x$Ga$_{1-x}$As waveguide. We further show that this technique is applicable to all $\chi^{(2)}$ material systems that utilize MPM for wavelength conversion. 
\end{abstract}


\section{Introduction}
Wavelength conversion in nonlinear materials  
plays an important role in a broad range of classical and quantum optical applications, including optical parametric amplification\cite{Lin_NC_2020}, supercontinuum generation\cite{Werner_OE_2019}, all-optical signal processing
\cite{Willner_JLT_14}, sensing\cite{Weichman_JMC_2019}, photon pair generation\cite{Torres_2011,Caspani_LSA_2017}, and squeezed state generation\cite{Dutt_PhysRevApplied_2015}, etc. Compact and efficient wavelength conversion can be achieved in nonlinear waveguides, in which the interacting fields are tightly confined for a long interaction length comparing to their bulk crystal counterparts. 
Waveguide based wavelength converters also make chip-scale integration with various active and passive components possible, enabling practical devices with new functionalities. In addition, waveguide dispersion provides an additional tuning knob in certain applications of wavelength conversion such as frequency comb generation\cite{Kues_NP_2019} and pure heralded single photon generation\cite{Kang_JOSAB_2014,Marchildon_Optica_2016}.  

Among popular $\chi^{(2)}$ nonlinear materials such as ferroelectrics (mainly lithium niobate), AlN, GaP, etc, compound semiconductor Al$_x$Ga$_{1-x}$As is uniquely attractive due to one of the highest $\chi^{(2)}$ coefficients ($d_{14}$$\approx$119 pm/V for GaAs), large transparency window from 0.9 $\mu$m to 17 $\mu$m, mature fabrication techniques, and perhaps in particular the ability of monolithic integration with active components due to its direct bandgap nature. However, phase matching in Al$_x$Ga$_{1-x}$As is challenging due to the formidable material dispersion near its bandgap and its non-ferroelectric nature. Several phase matching techniques have been developed for Al$_x$Ga$_{1-x}$As waveguides, including form-birefringence phase matching using low index AlOx layers\cite{Scaccabarozzi_OL_2006,Savanier_OE_2011}, quasi-phase matching (QPM) using etch and regrowth process \cite{Kuo_OL_2006,Yu_JCG_2007,Fedorava_OE_2013} or quantum-well intermixing (QWI) \cite{Helmy_OL_2000,Helmy_JAP_2006,Sarrafi_APL_2013}, modal phase matching (MPM) using either a higher order spatial mode\cite{Ducci_APL_2004,Duchesne_OE_2011} or a mode guided by Bragg reflections\cite{Abolghasem_PTL_2009,Abolghasem_JSTQE_2012}. 
More recently, QPM in suspended whisper-gallering-mode micro-disks\cite{Kuo_NC_2014} and snake-shaped curved waveguides\cite{Morais_OL_2017} have been demonstrated. In addition, high index contrast thin film AlGaAs-on-insulator waveguides enabled by wafer-bonding have shown the capabilities of birefringence phase matching\cite{May_OL_2019} as well as QPM in resonant micro-cavities\cite{Chang_APLPhontonics_2019}. 

While each of the above mentioned techniques shows certain advantages, the most efficient wavelength conversion to date, to the best of our knowledge, in any non-resonant monolithic platform was demonstrated in Bragg reflection waveguides (BRWs) employing MPM\cite{Abolghasem_PTL_2009}. Yet, among all phase matching techniques, MPM in general has the lowest modal field overlap due to the oscillating feature of one of the interacting fields. Although this limiting factor can be alleviated to some extent through epi-structure engineering which reduces the amount of negative electric field in the associated mode\cite{Ducci_APL_2004,Abolghasem_PTL_2009}, further improving the modal overlap, and thus the conversion efficiency, has not been reported in any Al$_x$Ga$_{1-x}$As waveguide using MPM.

In this work, we develop a technique to improve the conversion efficiency in a MPM waveguide by increasing its nonlinear modal overlap. Instead of attempting to optimize the shape of modal fields, we modify the material nonlinearity according to the sign of the modal electric field involved. Conversion efficiency enhancements are verified through a few design examples. Among them, this technique could improve the conversion efficiency by a factor of over 200 in the best case scenario. In addition, this technique is not limited to Al$_x$Ga$_{1-x}$As but can also be applied to all $\chi^{(2)}$ material systems which utilize MPM for wavelength conversion.  

\section{Theory and method}

We consider a generic three-wave mixing process, in which the pump (\textit{p}) and signal ($s$) beams are mixed to produce an idler (\textit{i}). For collinear propagation in the waveguide along the $z$-axis, the spatial-temporal dependence of the electric field of each interaction mode is given by
\begin{equation}
E_{j}(x,y,z,t)=A_j(z)E_j(x,y)\exp[-j(\beta_jz-\omega_jt)],
\label{Eq:field}
\end{equation}
where $j\in\{p,s,i\}$, and $A_j(z)$ and $E_j(x,y)$ are the slowly varying amplitude and spatial modal profile, respectively; $\omega_j$ is the angular frequency; $\beta_j$ is the propagation constant given by $\beta_j=2\pi n_j/\lambda_j$, where $n_j$ and $\lambda_j$ are the corresponding effective modal index and wavelength respectively. 

\par In the case of continuous wave excitation, we can neglect effects of group velocity mismatch and third-order nonlinearities. 
The couple-mode equations are given by\cite{Suhara_book_2003}  
\begin{equation}
\label{Eq:couple-mode pump}
\frac{\mathrm{d}A_p(z)}{\mathrm{d}z}=-j\rho_p\xi A_{s}^{\ast}A_i\exp[-j\Delta\beta z]-\frac{\alpha_p}{2}A_p,
\end{equation}
\begin{equation}
\label{Eq:couple-mode signal}
\frac{\mathrm{d}A_s(z)}{\mathrm{d}z}=-j\rho_s\xi A_{p}^{\ast}A_i\exp[-j\Delta\beta z]-\frac{\alpha_s}{2}A_s,
\end{equation}
\begin{equation}
\label{Eq:couple-mode idler}
\frac{\mathrm{d}A_i(z)}{\mathrm{d}z}=-j\rho_i\xi A_{p}A_s\exp[j\Delta\beta z]-\frac{\alpha_i}{2}A_i,
\end{equation}
where $\alpha_j$ is the linear loss coefficient and 
$\Delta\beta=\beta_i-\beta_s-\beta_p$ is the wave number mismatch. Here we define coefficients $\rho_j$ and the \textit{nonlinear overlap factor} $\xi$ as
\begin{equation}
\rho_j=(\frac{8\pi^2}{n_pn_sn_ic\epsilon_0\lambda_j^2})^{1/2},
\end{equation} 
\begin{equation} 
\xi=\frac{\iint{d(x,y)E_p^{\ast}(x,y)E_s(x,y)E_i(x,y)}\mathrm{d}x\mathrm{d}y}{(\iint|E_p^{\ast}(x,y)|^2\mathrm{d}x\mathrm{d}y\iint|E_s^{\ast}(x,y)|^2\mathrm{d}x\mathrm{d}y\iint|E_i^{\ast}(x,y)|^2\mathrm{d}x\mathrm{d}y)^{1/2}},
\label{Eq:overlap}
\end{equation} 
where $c$ and $\epsilon_0$ are the speed of light and vacuum permittivity, respectively, while $d(x,y)$ is the spatial distribution of the effective second order nonlinearity, after taking into account all tensor elements involved. In the case of $\omega_p=\omega_s$, Eqs. (\ref{Eq:couple-mode pump})--(\ref{Eq:couple-mode idler}) describe a second harmonic generation (SHG) process, otherwise a sum frequency generation (SFG) process. The coupled-mode equations for difference frequency generation (DFG) can be modified accordingly. 

In the absence of propagation losses, the normalized conversion efficiency, defined by $\eta=P_i/(P_pP_sL^2)$, is given by 
\begin{equation}
\eta=\rho_i^2\lvert\xi\rvert^2\sinc^2{(\Delta\beta L/2)},
\label{Eq:efficiency}
\end{equation}
where $L$ is the nonlinear waveguide length. According to Eq. (\ref{Eq:efficiency}), the maximum conversion efficiency in a lossless waveguide is uniquely determined by its nonlinear overlap factor given by Eq. (\ref{Eq:overlap}). 

\par
\begin{figure}[tbh!]
	\centering\includegraphics[width=0.8\textwidth]{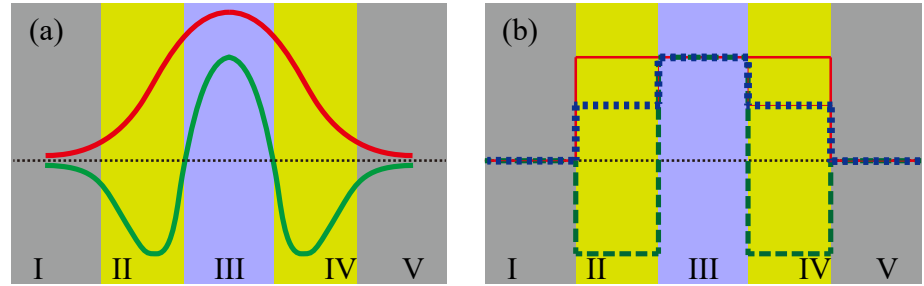}
	\caption{(a) Schematic of a three-layer slab waveguide using MPM with a fundamental mode (red) and a third order mode (green). The waveguide core consists of regions \Romannum{2}, \Romannum{3}, and \Romannum{4}, while the claddings are regions \Romannum{1} and \Romannum{5}. (b) Material nonlinearity distributions for the original waveguide (red solid line) and modified waveguides with negative nonlinearity (green dashed line) and reduced nonlinearity (blue dotted line). The claddings are assumed to have zero nonlinearity.}
	\label{fig:slab}
\end{figure}

 It should be noted that, the nonlinear overlap factor $\xi$ is not determined by the modal overlap of the interacting fields themselves, but instead the overlap of material nonlinearity and the modal fields involved. To illustrate this point, we show schematically in Fig. \ref{fig:slab} (a) a slab waveguide using MPM to achieve SHG, where two photons in a fundamental mode are converted to one photon of half the wavelength in a third order mode. Ideally, the two modes need to have the same spatial distribution to obtain the maximum conversion efficiency. However, the third order mode has both positive and negative electric field components, as shown in Fig. \ref{fig:slab} (a), which cancel out the contribution of each other in the overlap integral in Eq. (\ref{Eq:overlap}). This leads to a decreased nonlinear overlap factor and thus severely limits the conversion efficiency of MPM in general. 

\par To reduce this limitation, the conventional technique is to minimize the negative components of the modal field involved in MPM through epi-structure engineering, such as the third order mode in a optimized M-waveguide\cite{Ducci_APL_2004}, or the Bragg mode in a BRW\cite{Abolghasem_PTL_2009}. Nevertheless, the nonlinear overlap factors in such cases are still low, due to the unavoidable existence of negative field components. Here we propose an alternative solution which can overcome this limitation. Without even attempting to optimize the field profiles, instead, we modify the spatial distribution of the second order nonlinearity $d(x,y)$ such that it has negative values wherever the high order mode has negative modal components. As such, the negative parts of the modal field will not cancel out the positive parts in the overlap integral in Eq. (\ref{Eq:overlap}). 

\par Taking the structure shown in Fig. \ref{fig:slab} (a) for example, the nonlinear overlap factor, and thus the conversion efficiency, can be maximized if the material nonlinearity in 
regions \Romannum{2} and \Romannum{4} has a opposite sign to that of region \Romannum{3}, as shown in Fig. \ref{fig:slab} (b). This could be achieved by alternating the crystal orientation if the nonlinear coefficient is dependent on it, such as the case of compound semiconductors. If reversing the sign of nonlinearty is not achievable, the conversion efficiency could also be improved by reducing the magnitude of nonlinearty in regions \Romannum{2} and \Romannum{4}, or region \Romannum{3}, as also shown in Fig. \ref{fig:slab} (b). 

\section{Design examples}
\label{Sec. Design examples}

\subsection{Reversing the sign of $d(x,y)$}

\par
\begin{figure}[tbh!]
	\centering
	\includegraphics[width=0.9\textwidth]{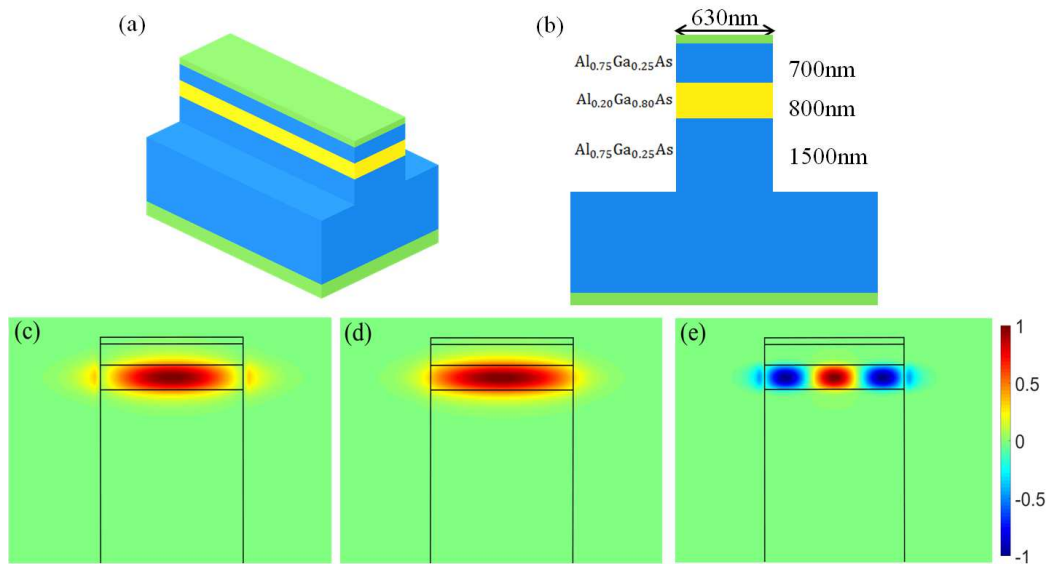}
	\caption{(a) Schematic of the Al$_x$Ga$_{1-x}$As ridge waveguide; (b) Waveguide cross-section with a ridge width of 630 nm for MPM; (c)-(e) The electric field profiles of the TE ($E_x$) and TM ($E_y$) fundamental modes at 1550 nm and that of the third-order TE$_{20}$ ($E_x$) mode at 775 nm, respectively.}
	\label{Fig:2D-AlGaAl}
\end{figure}

 To illustrate the technique presented above, we first provide an comparative example of Al$_x$Ga$_{1-x}$As waveguide utilizing MPM, as schematically shown in Fig. \ref{Fig:2D-AlGaAl}. The structure is grown on a GaAs [001]  substrate with a detailed epi-structure given in Fig. \ref{Fig:2D-AlGaAl}(b), and ridge waveguides are patterned along [110] direction. For a ridge width of 0.63 $\mu$m in a deeply etched waveguide, type-II phase matching is achieved between the TE$_{00}$ and TM$_{00}$ modes at 1550 nm and the TE$_{20}$ mode at 775 nm, with the modal electric field profiles shown in Figs. \ref{Fig:2D-AlGaAl}(c)-(e), respectively. The nonlinear overlap factor, and thus the conversion efficiency is limited the coexistence of both positive and negative electric field components in the second harmonic (SH).

\par
\begin{figure}[tb!]
	\centering
	\includegraphics[width=0.6\textwidth]{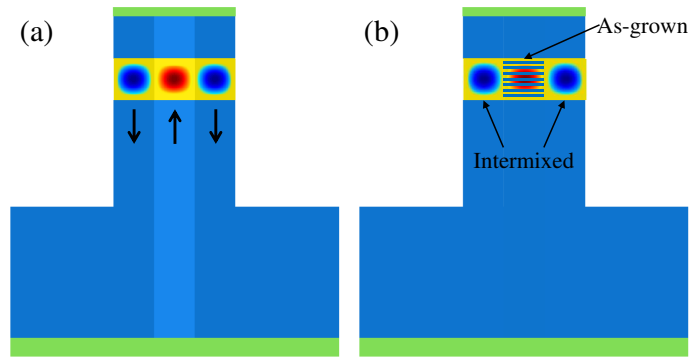}
	\caption{(a) Schematic of a domain inverted Al$_x$Ga$_{1-x}$As ridge waveguide superimposed with the electric field profile of the TE$_{20}$ mode at 775 nm, whose positive and negative components reside in domains of orthogonal orientations; (b) Schematics of an Al$_x$Ga$_{1-x}$As ridge waveguide employing QWI to selectively reduce the magnitude of nonlinearity in regions where the TE$_{20}$ mode at 775 nm has negative field components.}
	\label{Fig:AlGaAs modified designs}
\end{figure}

 A new design that achieves the maximal conversion efficiency requires the material nonlinearity $d(x,y)$ to be negative wherever the high order mode (TE$_{20}$ in this case) has negative electric field components. In Al$_x$Ga$_{1-x}$As, the effective value of $d(x,y)$ depends on the modal polarizations and the propagation direction. Its dependence on the propagation direction scales as $\cos{(2\theta)}$, with $\theta$ being the angle between propagation direction and the crystalline direction of [110]. As such, the sign of $d(x,y)$ is reversed if the waves propagate along  [$\bar{1}10$]. This property has been applied in a variety of QPM techniques, including orientation-patterned QPM (OP-QPM)\cite{Fedorava_OE_2013}, micro-ring and micro-disks\cite{Kuo_NC_2014}, as well as zig-zag\cite{Horn_arXiv_2010} and "snake-shaped" waveguides\cite{Morais_OL_2017}. By utilizing this property, we design a new structure with domain inversion, in which the crystal is rotated 90$^{\circ}$ in the regions where the TE$_{20}$ mode at SH has negative field components, as shown in Fig. \ref{Fig:AlGaAs modified designs}(a). Such waveguides could be fabricated by similar techniques used by orientation-patterned waveguides, except that the ridges are etched along, rather than perpendicular to the interfaces of alternating domains.

\par Since the material is optically isotropic, the modal profiles, effective indices and the phase matching condition in the new design are unchanged in the ideal case. With $d(x,y)$ having negative values where the TE$_{20}$ mode field components are also negative, the nonlinear overlap is increased by a factor of 17, leading to an increase of the conversion efficiency by a factor of $\sim$290 if the waveguide is lossless.

\par
\begin{figure}[tbh!]
	\centering
	\includegraphics[width=\textwidth]{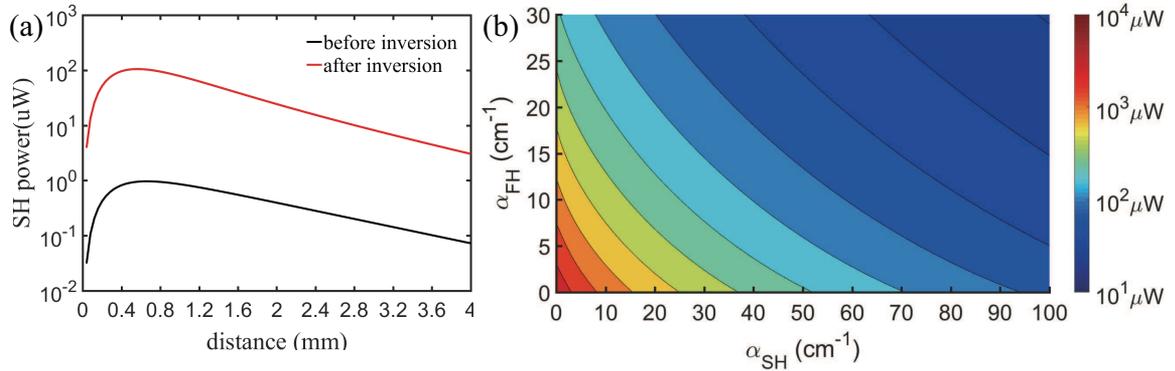}
	\caption{(a) SH output powers as functions of propagation distance in waveguides without and with domain inversion; (b) Maximal SH power as a function of fundamental frequency and SH propagation losses.}
	\label{Fig:Psh-vs-z}
\end{figure}

 In an actual waveguide, the conversion efficiency is also affected by propagation losses. We assumed typical losses of $\alpha_{\text{FH}}=4.2$ cm$^{-1}$ (18.3 dB/cm) for the fundamentals at 1550 nm and $\alpha_{\text{SH}}=73.7$ cm$^{-1}$ (320 dB/cm) for the SH\cite{Duchesne_OE_2011}, adopted from similar samples in the literatures, and solve the couple-mode equations given by Eqs. (\ref{Eq:couple-mode pump})--(\ref{Eq:couple-mode idler}) for SHG with an input power of 1 mW in each polarization. The dependence of SH powers on the propagation distance are given by Fig. \ref{Fig:Psh-vs-z}(a) for waveguides without and with domain inversion. Maximal SH powers of 0.97 $\mu$W and 105.1 $\mu$W are produced with optimal propagation distances of 0.68 mm and 0.56 mm, respectively, showing an increase of the normalized conversion efficiency $\eta$ by a factor of $\sim$160, from $5.2\times10^3$ \%W$^{-1}$cm$^{-2}$ to $8.4\times10^5$ \%W$^{-1}$cm$^{-2}$.

\par In the structure with domain inversion, the losses are likely higher than those in the reference design due to two additional interfaces. Therefore we investigated the dependence of maximal SH power on the losses, as shown in Fig. \ref{Fig:Psh-vs-z}(b). The maximal SH power decreases as the increase of fundamental and/or SH losses, as expected. However, an enhancement of the conversion efficiency could still be obtained as long as neither of the losses is increased by over 10 folds. As a comparison, the additional losses due to back scattering from hundreds of interfaces between inverted domains can be limited to a few dB/cm in the case of OP-QPM with the state of the art fabrication\cite{Yu_JCG_2007,Fedorava_OE_2013}. This value could be a reasonable expectation for the loss increase in our design.

\par Notice that fabricating ridge waveguides of such high aspect ratio and precision control is possible but challenging. In the above example, ridges as narrow as of $\sim$200 nm wide are required to be etched in the orientation-patterning process. This fabrication requirement can be eased significantly for longer phase matching wavelengths due to reduced material dispersions. For instance, a similar ridge waveguide can be designed to generate mid-infrared radiation at an output idler wavelength around 7.5 $\mu$m in its TE$_{00}$ mode via a DFG process with the input pump at around 1550 nm in its TE$_{20}$ mode and signal at 1950 nm in the TM$_{00}$ mode\cite{Logan_OL_2013}. The ridge width required for phase matching is increased by a factor of 3.5 to 2.2 $\mu$m. With domain inversion, the nonlinear overlap factor and the conversion efficiency are increased by a factor of 11 and 113, respectively, in a lossless waveguide. With representative loss values of 42.6 cm$^{-1}$ (185 dB/cm), 5.4 cm$^{-1}$ (23 dB/cm) and 18.6 cm$^{-1}$ (81 dB/cm) for pump, signal and idler, respectively, the normalized conversion efficiency is increased by a factor of 56 from 0.90 \%W$^{-1}$cm$^{-2}$ to 50.8 \%W$^{-1}$cm$^{-2}$.

\subsection{Reducing the magnitude of $d(x,y)$}

\par In cases that reversing the sign of $d(x,y)$ is impractical, the conversion efficiency could still be improved as long as the magnitude of $d(x,y)$ is reduced in the corresponding regions. In Al$_x$Ga$_{1-x}$As waveguides, the magnitude of $d(x,y)$ can be reduced by the technique of QWI if a quantum-well superlattice is embedded in the waveguide core\cite{Helmy_OL_2000,Helmy_JAP_2006}. This technique has been commonly used in domain-disordered QPM (DD-QPM), where the nonlinearity is periodically reduced along the direction of propagation. Comparing to OP-QPM, DD-QPM using QWI does not required any etch-and-regrowth fabrication process, and thus the waveguide losses are not considerably increased.

\par Here we consider QWI being used in previous design examples to reduce the magnitude of $d(x,y)$ instead of reversing its sign by orientation patterning. The superlattice core is selectively intermixed such that the magnitude of $d(x,y)$ is reduced in regions where the TE$_{20}$ mode has negative modal field components, as shown in Fig. \ref{Fig:AlGaAs modified designs}(b).  Assuming that $d(x,y)$ is reduced by a theoretical maximal value of 50 pm/V \cite{Hutchings_JSTQE_2004} while the waveguide losses are unchanged, the normalized conversion efficiencies for corresponding SHG and DFG processes are increased by factors of 4 and 9, to $1.9\times10^4$ \%W$^{-1}$cm$^{-2}$ and 7.8 \%W$^{-1}$cm$^{-2}$, respectively.

The realization of these structures using QWI in quantum confined heterostructures has been reported before by some of the authors. For example, having features on the order of 1 $\mu$m has been reported before, and the transition between intermixed and un-intermixed regions is on the sub-micron scale\cite{Helmy_JAP_2006}. After intermixing, plasma etching can be used to define the final width of the total waveguide width with an un-intermixed core and two lateral, intermixed layers on either side of that core. Nevertheless, the fabrication requirement can be lowered at a longer phase matching wavelength, where the material dispersion is smaller.

\par Note that this strategy of improving the nonlinear overlap factor does not only apply for MPM in compound semiconductors but also in all applicable material systems. The reduction of material nonlinearity can also be achieved by incorporating materials with smaller or no nonlinearity in a multi-layer waveguide core. Here we offer an example based on newly developed aluminum nitride (AlN)-on-insulator platform, which has been demonstrated for applications such as electro-optic modulation\cite{Xiong_NL_2012}, nonlinear wavelength conversion\cite{Guo_Optica_2016}, as well as photon pair generation\cite{Guo_LSA_2017}. MPM between the TM$_{00}$ mode at 1550 nm and the TM$_{20}$ mode at 775 nm in AlN waveguides can be achieved by choosing an appropriate waveguide dimension, similar to previously shown Al$_x$Ga$_{1-x}$As waveguide designs. The c axis of polycrystalline AlN is perpendicular to the wafer in order to exploit the largest component of AlN's $\chi^{(2)}$ tensor ($d_{33}$)\cite{Xiong_NL_2012}.

\begin{figure}[tbh!]
	\centering
	\includegraphics[width=0.4\textwidth]{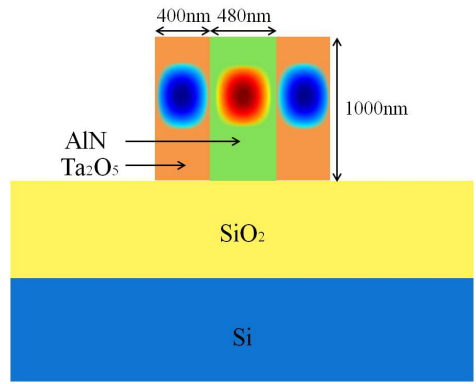}
	\caption{Schematic of the modified AlN-on-insulator ridge waveguide superimposed with field profile of the TM$_{20}$ mode at 775 nm.}
	\label{Fig:AlN}
\end{figure}

\par In the modified design, the AlN ridge is sandwiched by a pair of Ta$_2$O$_5$ dielectric such that the positive part of the TM$_{20}$ mode at 775 nm locates in AlN while the negative parts locate in Ta$_2$O$_5$, as shown in Fig. \ref{Fig:AlN}. The fabrication of these structures can be easily achieved using a two-step lithography process, where initial core ridges are fabricated using lithography and plasma etching. A subsequent process of conformal deposition of the outer regions of the core is carried out. The deposition time and conformal profile of this process can be tuned to suite the required dimensions of the outer layers of the core. Finally, a lithography along with plasma etching process can then follow to achieve the profile and dimensions discussed in this work. 
Here Ta$_2$O$_5$ is chosen because it has an index of refraction close to that of AlN, and it has near-zero second order nonlinearity. This leads to an increase of the nonlinear overlap factor by a factor of $\sim$12 and the SHG efficiency by 144 without waveguide losses. With realistic losses of 2.6 cm$^{-1}$ (11.3 dB/cm) and 86 cm$^{-1}$ (373 dB/cm) for the fundamental and SH\cite{Xiong_NL_2012}, the normalized efficiency is increased by a factor of 12 from 0.11 \%W$^{-1}$cm$^{-2}$ to 1.27 \%W$^{-1}$cm$^{-2}$.

\section{Discussion and conclusion}

\par Design examples provided above demonstrate the effectiveness of the proposed technique. In the best case scenario, we observe up to $\sim$290 folds enhancement in the conversion efficiency assuming the waveguide does not include any losses. With the existence of losses, the conversion efficiency can still be improved significantly. The results are summarized in Table \ref{table}. It should be emphasized that the technique proposed here applies to $\chi^{(2)}$ nonlinear waveguides of all material systems that adopt MPM in principle, despite it was mainly motivated for compound semiconductors.
\begin{table}[tbh!]
	\centering
	\caption{Summary of normalized conversion efficiencies for all design examples provided in Sec. \ref{Sec. Design examples}. Here $\eta_{\text{ref}}$ represents the normalized conversion efficiency in the reference design, while $\eta_{\text{mod}}$ represents that of the modified design using the proposed technique. Waveguide losses are all included in the calculations.}
	\label{table}
	\begin{tabular}{ccccc}
		\hline
		example                                                                        & \begin{tabular}[c]{@{}c@{}}frequency\\ conversion process\end{tabular} & \begin{tabular}[c]{@{}c@{}}$\eta_{\text{ref}}$\\  (\%W$^{-1}$cm$^{-2}$)\end{tabular} & \begin{tabular}[c]{@{}c@{}}$\eta_{\text{mod}}$ \\ (\%W$^{-1}$cm$^{-2}$)\end{tabular} & $\eta_{\text{mod}}/\eta_{\text{ref}}$ \\ \hline
		\multirow{2}{*}{\begin{tabular}[c]{@{}c@{}}Al$_x$Ga$_{1-x}$As\\ inversion\end{tabular}} & $\omega_{1550}+\omega_{1550}\rightarrow\omega_{775}$                   & $5.2\times10^3$                                                                      & $8.4\times10^5$                                                                      & 160                                           \\ \cline{2-5} 
		& $\omega_{1550}-\omega_{1950}\rightarrow\omega_{7556}$                  & 0.90                                                                                 & 50.8                                                                                 & 56                                            \\ \hline
		\multirow{2}{*}{\begin{tabular}[c]{@{}c@{}}Al$_x$Ga$_{1-x}$As\\ QWI\end{tabular}}       & $\omega_{1550}+\omega_{1550}\rightarrow\omega_{775}$                   & $5.2\times10^3$                                                                      & $1.9\times10^4$                                                                      & 4                                             \\ \cline{2-5} 
		& $\omega_{1550}-\omega_{1950}\rightarrow\omega_{7556}$                  & 0.90                                                                                 & 7.8                                                                                  & 9                                             \\ \hline
		AlN                                                                            & $\omega_{1550}+\omega_{1550}\rightarrow\omega_{775}$                   & 0.11                                                                                 & 1.27                                                                                 & 12                                            \\ \hline
	\end{tabular}
\end{table} 


\par The use of specific spatial modes in these designs poses strict constraints on the waveguide width in each etch process. Deviation from the ideal value causes the shift of phase matching wavelength and the decrease of nonlinear overlap. Fabrication of these structures is certainly nontrivial but still practical, as discussed above. After all, techniques including orientation-patterning and QWI have been developed in domain-reversal and domain-suppressed QPM respectively, and high aspect ratio waveguides with low losses and precision etch control have also been achieved\cite{Porkolab_OE_2014,Liao_OME_2017}. In addition, the use of a higher order mode may first seem intimidating as it often has no easy access. However, waveguide couplers can be implemented to selectively couple light from a neighboring single mode waveguide into the higher order mode, and vice versa, as demonstrated in \cite{Guo_Optica_2016,Guo_LSA_2017}.

\par 
Although we use TE$_{20}$ and TM$_{20}$ modes in the design examples, this technique could be applied to MPM waveguides using any other spatial mode	provided the material nonlinearity can be modified accordingly. In above designs, material nonlinearities are altered horizontally (parallel to the wafer surface), and horizontal higher order modes (TE$_{20}$ and TM$_{20}$) are defined lithographically to achieve phase matching. Instead, if material nonlinearities can be modified in the vertical direction (perpendicular to the wafer surface), and the thickness of each layer can be accurately controlled, vertical higher order modes, such as TE$_{02}$ and TM$_{02}$, can be used. In such cases, requirements on ridge width and etch depth can be significantly relaxed. For instance, in Al$_x$Ga$_{1-x}$As, quantum-well superlattices could be embedded in the epi-stricture of a M-shaped waveguide\cite{Ducci_APL_2004} or a BRW\cite{Abolghasem_PTL_2009,Abolghasem_JSTQE_2012,Logan_OL_2013} followed by intermixing to reduce the material nonlinearity in certain regions. As for AlN, it can be deposited on a layer of Ta$_2$O$_5$ on silica, followed by a cap Ta$_2$O$_5$ layer. In this case, the TM$_{02}$ mode is used for the SH, while the ridge width is non-essential. 
During the preparation of this manuscript, we noticed a recent work based on thin film lithium niobate waveguides which took a strategy similar to this AlN example to increase the conversion efficiency\cite{Luo_LPR_2019}.

\par This technique could be used in conjunction with resonant structures to further enhance the conversion efficiency. For example, the Al$_x$Ga$_{1-x}$As waveguides given above can be embedded in Fabry-P\'erot type cavities, and the AlN waveguides can be lithographically defined as ring resonators\cite{Guo_Optica_2016}.

\par In summary, we have developed a technique to increase the nonlinear overlap factor and therefore the wavelength conversion efficiency in $\chi^{(2)}$ waveguides utilizing MPM. In contrary to the conventional method of optimizing modal field overlap, we modify the material nonlinearity by either reversing the sign or reducing the magnitude in regions which coincide with the negative parts of the high frequency mode. This could eliminate the limiting factor of low nonlinear overlap inherent to MPM and thus fully harness the material nonlinearity for efficient wavelength conversions.



\bibliographystyle{IEEEtran}
\bibliography{reference}

\begin{thebibliography}{10}
\providecommand{\url}[1]{#1}
\csname url@samestyle\endcsname
\providecommand{\newblock}{\relax}
\providecommand{\bibinfo}[2]{#2}
\providecommand{\BIBentrySTDinterwordspacing}{\spaceskip=0pt\relax}
\providecommand{\BIBentryALTinterwordstretchfactor}{4}
\providecommand{\BIBentryALTinterwordspacing}{\spaceskip=\fontdimen2\font plus
\BIBentryALTinterwordstretchfactor\fontdimen3\font minus
  \fontdimen4\font\relax}
\providecommand{\BIBforeignlanguage}[2]{{%
\expandafter\ifx\csname l@#1\endcsname\relax
\typeout{** WARNING: IEEEtran.bst: No hyphenation pattern has been}%
\typeout{** loaded for the language `#1'. Using the pattern for}%
\typeout{** the default language instead.}%
\else
\language=\csname l@#1\endcsname
\fi
#2}}
\providecommand{\BIBdecl}{\relax}
\BIBdecl

\bibitem{Lin_NC_2020}
Y.~Lin, Y.~Nabekawa, and K.~Midorikawa, ``Optical parametric amplification of
  sub-cycle shortwave infrared pulses,'' \emph{Nature Communications}, vol.~11,
  p. 3413, 2020.

\bibitem{Werner_OE_2019}
\BIBentryALTinterwordspacing
K.~Werner, M.~G. Hastings, A.~Schweinsberg, B.~L. Wilmer, D.~Austin, C.~M.
  Wolfe, M.~Kolesik, T.~R. Ensley, L.~Vanderhoef, A.~Valenzuela, and
  E.~Chowdhury, ``Ultrafast mid-infrared high harmonic and supercontinuum
  generation with $n_2$ characterization in zinc selenide,'' \emph{Opt.
  Express}, vol.~27, no.~3, pp. 2867--2885, Feb 2019. [Online]. Available:
  \url{http://www.opticsexpress.org/abstract.cfm?URI=oe-27-3-2867}
\BIBentrySTDinterwordspacing

\bibitem{Willner_JLT_14}
\BIBentryALTinterwordspacing
A.~E. Willner, S.~Khaleghi, M.~R. Chitgarha, and O.~F. Yilmaz, ``All-optical
  signal processing,'' \emph{J. Lightwave Technol.}, vol.~32, no.~4, pp.
  660--680, Feb 2014. [Online]. Available:
  \url{http://jlt.osa.org/abstract.cfm?URI=jlt-32-4-660}
\BIBentrySTDinterwordspacing

\bibitem{Weichman_JMC_2019}
\BIBentryALTinterwordspacing
M.~L. Weichman, P.~B. Changala, J.~Ye, Z.~Chen, M.~Yan, and N.~Picqu\'e,
  ``Broadband molecular spectroscopy with optical frequency combs,''
  \emph{Journal of Molecular Spectroscopy}, vol. 355, pp. 66--78, 2019.
  [Online]. Available:
  \url{http://www.sciencedirect.com/science/article/pii/S0022285218302881}
\BIBentrySTDinterwordspacing

\bibitem{Torres_2011}
\BIBentryALTinterwordspacing
J.~P. Torres, K.~Banaszek, and I.~Walmsley, ``Chapter 5 - engineering nonlinear
  optic sources of photonic entanglement,'' ser. Progress in Optics, E.~Wolf,
  Ed.\hskip 1em plus 0.5em minus 0.4em\relax Elsevier, 2011, vol.~56, pp.
  227--331. [Online]. Available:
  \url{http://www.sciencedirect.com/science/article/pii/B9780444538864000058}
\BIBentrySTDinterwordspacing

\bibitem{Caspani_LSA_2017}
L.~Caspani, C.~Xiong, B.~J. Eggleton, D.~Bajoni, M.~Liscidini, M.~Galli,
  R.~Morandotti, and D.~J. Moss, ``Integrated sources of photon quantum states
  based on nonlinear optics,'' \emph{Light: Science \& Applications}, vol.~6,
  p. e17100, 2017.

\bibitem{Dutt_PhysRevApplied_2015}
\BIBentryALTinterwordspacing
A.~Dutt, K.~Luke, S.~Manipatruni, A.~L. Gaeta, P.~Nussenzveig, and M.~Lipson,
  ``On-chip optical squeezing,'' \emph{Phys. Rev. Applied}, vol.~3, p. 044005,
  Apr 2015. [Online]. Available:
  \url{https://link.aps.org/doi/10.1103/PhysRevApplied.3.044005}
\BIBentrySTDinterwordspacing

\bibitem{Kues_NP_2019}
M.~Kues, C.~Reimer, J.~M. Lukens, W.~J. Munro, A.~M. Weiner, D.~J. Moss, and
  R.~Morandotti, ``Quantum optical microcombs,'' \emph{Nature Photonics},
  vol.~13, pp. 170--179, 2019.

\bibitem{Kang_JOSAB_2014}
\BIBentryALTinterwordspacing
D.~Kang, A.~Pang, Y.~Zhao, and A.~S. Helmy, ``Two-photon quantum state
  engineering in nonlinear photonic nanowires,'' \emph{J. Opt. Soc. Am. B},
  vol.~31, no.~7, pp. 1581--1589, Jul 2014. [Online]. Available:
  \url{http://josab.osa.org/abstract.cfm?URI=josab-31-7-1581}
\BIBentrySTDinterwordspacing

\bibitem{Marchildon_Optica_2016}
\BIBentryALTinterwordspacing
R.~P. Marchildon and A.~S. Helmy, ``Dispersion-enabled quantum state control in
  integrated photonics,'' \emph{Optica}, vol.~3, no.~3, pp. 243--251, Mar 2016.
  [Online]. Available:
  \url{http://www.osapublishing.org/optica/abstract.cfm?URI=optica-3-3-243}
\BIBentrySTDinterwordspacing

\bibitem{Scaccabarozzi_OL_2006}
\BIBentryALTinterwordspacing
L.~Scaccabarozzi, M.~M. Fejer, Y.~Huo, S.~Fan, X.~Yu, and J.~S. Harris,
  ``Enhanced second-harmonic generation in {AlGaAs/AlxOy} tightly confining
  waveguides and resonant cavities,'' \emph{Opt. Lett.}, vol.~31, no.~24, pp.
  3626--3628, Dec 2006. [Online]. Available:
  \url{http://ol.osa.org/abstract.cfm?URI=ol-31-24-3626}
\BIBentrySTDinterwordspacing

\bibitem{Savanier_OE_2011}
\BIBentryALTinterwordspacing
M.~Savanier, A.~Andronico, A.~Lema\^{i}tre, C.~Manquest, I.~Favero, S.~Ducci,
  and G.~Leo, ``Nearly-degenerate three-wave mixing at 1.55 $\mu$m in oxidized
  {AlGaAs} waveguides,'' \emph{Opt. Express}, vol.~19, no.~23, pp.
  22\,582--22\,587, Nov 2011. [Online]. Available:
  \url{http://www.opticsexpress.org/abstract.cfm?URI=oe-19-23-22582}
\BIBentrySTDinterwordspacing

\bibitem{Kuo_OL_2006}
\BIBentryALTinterwordspacing
P.~S. Kuo, K.~L. Vodopyanov, M.~M. Fejer, D.~M. Simanovskii, X.~Yu, J.~S.
  Harris, D.~Bliss, and D.~Weyburne, ``Optical parametric generation of a
  mid-infrared continuum in orientation-patterned {GaAs},'' \emph{Opt. Lett.},
  vol.~31, no.~1, pp. 71--73, Jan 2006. [Online]. Available:
  \url{http://ol.osa.org/abstract.cfm?URI=ol-31-1-71}
\BIBentrySTDinterwordspacing

\bibitem{Yu_JCG_2007}
\BIBentryALTinterwordspacing
X.~Yu, L.~Scaccabarozzi, A.~C. Lin, M.~M. Fejer, and J.~S. Harris, ``Growth of
  {GaAs} with orientation-patterned structures for nonlinear optics,''
  \emph{Journal of Crystal Growth}, vol. 301-302, no.~0, pp. 163 -- 167, 2007.
  [Online]. Available:
  \url{http://www.sciencedirect.com/science/article/pii/S0022024806014345}
\BIBentrySTDinterwordspacing

\bibitem{Fedorava_OE_2013}
K.~Fedorova, A.~McRobbie, G.~Sokolovskii, P.~Schunemann, and E.~Rafailov,
  ``Second harmonic generation in a low-loss orientation-patterned {GaAs}
  waveguide,'' \emph{Opt. Express}, vol.~21, no.~14, pp. 16\,424--16\,430,
  2013.

\bibitem{Helmy_OL_2000}
\BIBentryALTinterwordspacing
A.~S. Helmy, D.~C. Hutchings, T.~C. Kleckner, J.~H. Marsh, A.~C. Bryce, J.~M.
  Arnold, C.~R. Stanley, J.~S. Aitchison, C.~T.~A. Brown, K.~Moutzouris, and
  M.~Ebrahimzadeh, ``Quasi phase matching in {GaAs-AlAs} superlattice
  waveguides through bandgap tuning by use of quantum-well intermixing,''
  \emph{Opt. Lett.}, vol.~25, no.~18, pp. 1370--1372, Sep 2000. [Online].
  Available: \url{http://ol.osa.org/abstract.cfm?URI=ol-25-18-1370}
\BIBentrySTDinterwordspacing

\bibitem{Helmy_JAP_2006}
\BIBentryALTinterwordspacing
A.~S. Helmy, A.~C. Bryce, D.~C. Hutchings, J.~S. Aitchison, and J.~H. Marsh,
  ``Band gap gratings using quantum well intermixing for
  quasi-phase-matching,'' \emph{Journal of Applied Physics}, vol. 100, no.~12,
  p. 123107, 2006. [Online]. Available: \url{https://doi.org/10.1063/1.2402034}
\BIBentrySTDinterwordspacing

\bibitem{Sarrafi_APL_2013}
P.~Sarrafi, E.~Y. Zhu, K.~Dolgaleva, B.~M. Holmes, D.~C. Hutchings, J.~S.
  Aitchison, and L.~Qian, ``Continuous-wave quasi-phase-matched waveguide
  correlated photon pair source on a {III--V} chip,'' \emph{Applied Physics
  Letters}, vol. 103, no.~25, p. 251115, 2013.

\bibitem{Ducci_APL_2004}
\BIBentryALTinterwordspacing
S.~Ducci, L.~Lanco, V.~Berger, A.~D. Rossi, V.~Ortiz, and M.~Calligaro,
  ``Continuous-wave second-harmonic generation in modal phase matched
  semiconductor waveguides,'' \emph{Applied Physics Letters}, vol.~84, no.~16,
  pp. 2974--2976, 2004. [Online]. Available:
  \url{http://link.aip.org/link/?APL/84/2974/1}
\BIBentrySTDinterwordspacing

\bibitem{Duchesne_OE_2011}
\BIBentryALTinterwordspacing
D.~Duchesne, K.~A. Rutkowska, M.~Volatier, F.~L\'{e}gar\'{e}, S.~Delprat,
  M.~Chaker, D.~Modotto, A.~Locatelli, C.~D. Angelis, M.~Sorel, D.~N.
  Christodoulides, G.~Salamo, R.~Ar\`{e}s, V.~Aimez, and R.~Morandotti,
  ``Second harmonic generation in {AlGaAs} photonic wires using low power
  continuous wave light,'' \emph{Opt. Express}, vol.~19, no.~13, pp.
  12\,408--12\,417, Jun 2011. [Online]. Available:
  \url{http://www.opticsexpress.org/abstract.cfm?URI=oe-19-13-12408}
\BIBentrySTDinterwordspacing

\bibitem{Abolghasem_PTL_2009}
P.~Abolghasem, J.~Han, B.~J. Bijlani, A.~Arjmand, and A.~S. Helmy, ``Highly
  efficient second-harmonic generation in monolithic matching layer enhanced
  {Al$_x$Ga$_{1-x}$As} waveguides,'' \emph{Photonics Technology Letters, IEEE},
  vol.~21, no.~19, pp. 1462--1464, 2009.

\bibitem{Abolghasem_JSTQE_2012}
P.~Abolghasem, J.-B. Han, D.~Kang, B.~J. Bijlani, and A.~S. Helmy, ``Monolithic
  photonics using second-order optical nonlinearities in multilayer-core
  {Bragg} reflection waveguides,'' \emph{Selected Topics in Quantum
  Electronics, IEEE Journal of}, vol.~18, no.~2, pp. 812--825, 2012.

\bibitem{Kuo_NC_2014}
P.~S. Kuo, J.~Bravo-Abad, and G.~S. Solomon, ``Second-harmonic generation using
  quasi-phasematching in a {GaAs} whispering-gallery-mode microcavity,''
  \emph{Nature Communications}, vol.~5, p. 3109, 2014.

\bibitem{Morais_OL_2017}
\BIBentryALTinterwordspacing
N.~Morais, I.~Roland, M.~Ravaro, W.~Hease, A.~Lema\^{i}tre, C.~Gomez,
  S.~Wabnitz, M.~D. Rosa, I.~Favero, and G.~Leo, ``Directionally induced
  quasi-phase matching in homogeneous {AlGaAs} waveguides,'' \emph{Opt. Lett.},
  vol.~42, no.~21, pp. 4287--4290, Nov 2017. [Online]. Available:
  \url{http://ol.osa.org/abstract.cfm?URI=ol-42-21-4287}
\BIBentrySTDinterwordspacing

\bibitem{May_OL_2019}
\BIBentryALTinterwordspacing
S.~May, M.~Kues, M.~Clerici, and M.~Sorel, ``Second-harmonic generation in
  {AlGaAs}-on-insulator waveguides,'' \emph{Opt. Lett.}, vol.~44, no.~6, pp.
  1339--1342, Mar 2019. [Online]. Available:
  \url{http://ol.osa.org/abstract.cfm?URI=ol-44-6-1339}
\BIBentrySTDinterwordspacing

\bibitem{Chang_APLPhontonics_2019}
\BIBentryALTinterwordspacing
L.~Chang, A.~Boes, P.~Pintus, J.~D. Peters, M.~Kennedy, X.-W. Guo, N.~Volet,
  S.-P. Yu, S.~B. Papp, and J.~E. Bowers, ``Strong frequency conversion in
  heterogeneously integrated {GaAs} resonators,'' \emph{APL Photonics}, vol.~4,
  no.~3, p. 036103, 2019. [Online]. Available:
  \url{https://doi.org/10.1063/1.5065533}
\BIBentrySTDinterwordspacing

\bibitem{Suhara_book_2003}
T.~Suhara and M.~Fujimura, \emph{Waveguide nonlinear-optic devices}.\hskip 1em
  plus 0.5em minus 0.4em\relax Springer, 2003.

\bibitem{Horn_arXiv_2010}
R.~T. Horn and G.~Weihs, ``Directional quasi-phase matching in curved
  waveguides,'' \emph{arXiv preprint arXiv:1008.2190}, 2010.

\bibitem{Logan_OL_2013}
\BIBentryALTinterwordspacing
D.~F. Logan, M.~Giguere, A.~Villeneuve, and A.~S. Helmy, ``Widely tunable
  mid-infrared generation via frequency conversion in semiconductor
  waveguides,'' \emph{Opt. Lett.}, vol.~38, no.~21, pp. 4457--4460, Nov 2013.
  [Online]. Available: \url{http://ol.osa.org/abstract.cfm?URI=ol-38-21-4457}
\BIBentrySTDinterwordspacing

\bibitem{Hutchings_JSTQE_2004}
D.~C. {Hutchings}, ``Theory of ultrafast nonlinear refraction in semiconductor
  superlattices,'' \emph{IEEE Journal of Selected Topics in Quantum
  Electronics}, vol.~10, no.~5, pp. 1124--1132, Sep. 2004.

\bibitem{Xiong_NL_2012}
\BIBentryALTinterwordspacing
C.~Xiong, W.~H.~P. Pernice, and H.~X. Tang, ``Low-loss, silicon integrated,
  aluminum nitride photonic circuits and their use for electro-optic signal
  processing,'' \emph{Nano Letters}, vol.~12, no.~7, pp. 3562--3568, 2012,
  pMID: 22663299. [Online]. Available: \url{https://doi.org/10.1021/nl3011885}
\BIBentrySTDinterwordspacing

\bibitem{Guo_Optica_2016}
X.~Guo, C.-L. Zou, and H.~X. Tang, ``Second-harmonic generation in aluminum
  nitride microrings with 2500\%/{W} conversion efficiency,'' \emph{Optica},
  vol.~3, no.~10, pp. 1126--1131, Oct. 2016.

\bibitem{Guo_LSA_2017}
X.~Guo, C.-L. Zou, C.~Schuck, H.~Jung, and H.~X. Tang, ``Parametric
  down-conversion photon-pair source on a nanophotonic chip,'' \emph{Light:
  Science \& Applications}, vol.~6, p. e16249, May 2017.

\bibitem{Porkolab_OE_2014}
\BIBentryALTinterwordspacing
G.~A. Porkolab, P.~Apiratikul, B.~Wang, S.~H. Guo, and C.~J.~K. Richardson,
  ``Low propagation loss {AlGaAs} waveguides fabricated with plasma-assisted
  photoresist reflow,'' \emph{Opt. Express}, vol.~22, no.~7, pp. 7733--7743,
  Apr 2014. [Online]. Available:
  \url{http://www.opticsexpress.org/abstract.cfm?URI=oe-22-7-7733}
\BIBentrySTDinterwordspacing

\bibitem{Liao_OME_2017}
\BIBentryALTinterwordspacing
Z.~Liao and J.~S. Aitchison, ``Precision etching for multi-level {AlGaAs}
  waveguides,'' \emph{Opt. Mater. Express}, vol.~7, no.~3, pp. 895--903, Mar
  2017. [Online]. Available:
  \url{http://www.osapublishing.org/ome/abstract.cfm?URI=ome-7-3-895}
\BIBentrySTDinterwordspacing

\bibitem{Luo_LPR_2019}
\BIBentryALTinterwordspacing
R.~Luo, Y.~He, H.~Liang, M.~Li, and Q.~Lin, ``Semi-nonlinear nanophotonic
  waveguides for highly efficient second-harmonic generation,'' \emph{Laser \&
  Photonics Reviews}, vol.~13, no.~3, p. 1800288, 2019. [Online]. Available:
  \url{https://onlinelibrary.wiley.com/doi/abs/10.1002/lpor.201800288}
\BIBentrySTDinterwordspacing

\end{thebibliography}



\end{document}